\documentclass[12pt]{article}
\usepackage{graphicx}

\newcounter{obs}
\begin{document}

\centerline{\bf THE ANGULAR MOMENTUM-ENERGY SPACE}

 \vspace{7mm}

\centerline{Dan Com\u anescu}

\vspace{10mm}

{\footnotesize

 \noindent {\bf Abstract.} In this paper we shall define and study the angular momentum-energy space for the classical problem of plane-motions of a particle
 situated in a potential field of a central force. We shall present the angular momentum-energy space for some important cases.

 \vspace{3mm}


\vspace{2mm}


}

\vspace{10mm}

\centerline{\bf 1. Introduction}

\vspace{5mm}

The angular momentum-energy states are used in Classical Mechanics to construct a mathematical model for the plane motions of a particle in a
potential field of a central force (see [1], [4] or [5]).

In Astrophysics appears an equation in the angular momentum-energy space describing the stellar distribution around a black hole (see [2]).

In General Relativity a mathematical model of the motions of a particle use the concepts of energy and angular momentum (see [3]).
\vspace{2mm}

The objectives of this study are:

\begin{itemize}
    \item to present the classical concepts of angular momentum and energy;
    \item to study the Angular Momentum-Energy Space and the Angular Momentum-Energy Space which are corresponding to the uniform
    rotations;
    \item to present the Angular Momentum-Energy Space and Angular Momentum-Energy Space which are corresponding to the uniform
    rotations for some particular force fields.
\end{itemize}
\vspace{10mm}

\centerline{\bf 2. The Movements of a Particle in a Potential Field of a Central Force}

\vspace{5mm}

We consider a particle situated in a potential field of central force. In the Newtonian Mechanics it is known that a trajectory is contained in
a plane which contains the center of the force. We study the case in which the trajectories are contained in a fixed plane passing through
\emph{center} $O.$ We denote by $\overrightarrow{r}$ \emph{the radius vector} of the particle, $r$ \emph{the modulus of the radius vector} and
$U_{\ast }(r)$ \emph{the force function}. In this case the second law of Newton has the form:
 \begin{equation}\label{*}
 m\ddot{\overrightarrow{r}}=-U'_{*}(r)\frac{\overrightarrow{r}}{r}
\end{equation}
Projecting this equation on the natural base of the \emph{polar coordinates} $(r,\varphi )$ \ we have:
 \begin{equation}\label{*}
    \left\{%
\begin{array}{ll}
    \dot{\overbrace{r^{2}\dot{\varphi}}}=0 \\
    \ddot{r}-r\dot{\varphi}^{2}+U'(r)=0\\
\end{array}%
\right.
\end{equation}
where $U(r)=U_{\ast }(r)/m$ is \emph{the force function per unit mass}.

We denote by $J_{\ast }$ \emph{the angular momentum}, $E_{\ast }$ \emph{the total energy}, $J=J_{\ast }/m$ \emph{the angular momentum per unit
mass} and $E=E_{\ast }/m$ \emph{the total energy per unit mass}.

An other mathematical model of the motions is obtained using the conservation laws of the angular momentum and total energy. We have:
\begin{equation}\label{*}
    \left\{%
\begin{array}{ll}
    r^{2}\dot{\varphi}=J \\
    \frac{\dot{r}^{2}}{2}+\frac{J^{2}}{2r^{2}}+U(r)=E\\
\end{array}%
\right.
\end{equation}
\vspace{2mm}

\noindent \textbf{Theorem 2.1.} i) \emph{If $(r,\varphi )$ is a solution of (2) and it is not an uniform rotation, then it exists $(J,E)\in
\textbf{R}^{2}$ such that $(r,\varphi )$ is a solution of (3).}

\noindent ii) \emph{If $(J,E)\in \textbf{R}^{2}$, $(r,\varphi )$ is a solution of (3) and it is not an uniform rotation, then $(r,\varphi )$ is
a solution of (2).}

\noindent iii) \emph{If $r_{0}>0$, then it exists an uniform rotation $(r_{0},\varphi )$ solution of (2) and (3) if and only if:}
\begin{equation}\label{*}
    \left\{%
\begin{array}{ll}
    J^{2}=r_{0}^{3}U'(r_{0}) \\
    E=U(r_{0})+\frac{r_{0}U'(r_{0})}{2}\\
\end{array}%
\right.
\end{equation}
\vspace{2mm}

\noindent \emph{Proof.} The propositions i) and ii) are classical results.

\noindent iii) Let $(J,E)\in \textbf{R}^{2}$ and $r_{0}>0$ such that the relation (4) is true. The uniform rotation
$(r_{0},\frac{Jt}{r_{0}^{2}})$ is a solution of (2) and (3). Let $(J,E)\in \textbf{R}^{2}$, $r_{0}>0$ and $(r_{0},\varphi )$ an uniform rotation
which is a solution of (2) and (3) then we have:
\begin{equation}\label{*}
\dot{\varphi}=\frac{J}{r_{0}^{2}},\,\,\,r_{0}\dot{\varphi}^{2}-U'(r_{0})=0,\,\,\,\frac{J^{2}}{2r_{0}^{2}}+U(r_{0})=E.
\end{equation}
We introduce (5)$_{1}$ in (5)$_{2}$ and we obtain (4)$_{1}.$ The relation (4)$_{2}$ is obtained using (4)$_{1}$ and (5)$_{3}$. \vspace{2mm}

\noindent \textbf{Hypothesis}: \emph{In this paper we suppose that a force function per unit mass is a function} $U\in C^{1}((0,\infty
),\textbf{R})$. \vspace{2mm}

\noindent \textbf{Remark 2.1.} The most important is the case of an attractive force field which is characterized by a force function with the
property $U'>0$. \vspace{2mm}

In this paper we use the following notations: \emph{the effective force function per unit mass}:
\begin{equation}\label{*}
\textbf{V}_{J}^{U}(r)=\frac{J^{2}}{2r^{2}}+U(r)
\end{equation}
\emph{the effective angular momentum per unit mass}:
\begin{equation}\label{*}
\textbf{W}_{E}^{U}(r)=2r^{2}(E-U(r))
\end{equation}
where $U$ is a force function per unit mass, $E$ the total energy per unit mass and $J$ the angular momentum per unit mass. We have
$\textbf{V}_{J}^{U},\textbf{W}_{E}^{U}\in C^{1}((0,\infty ),\textbf{R}).$

It is easy to see: \vspace{2mm}

\noindent \textbf{Proposition 2.1.} \emph{If $(r,\varphi )$ is a solution of (3), then for all time-moments we have:}
\begin{equation}\label{*}
\textbf{V}_{J}^{U}(r(t))\leq E
\end{equation}
\begin{equation}\label{*}
\textbf{W}_{E}^{U}(r(t))\geq J^{2}
\end{equation}
\vspace{2mm}

The most important notions for the paper are presented in the next considerations. \vspace{2mm}

\noindent \textbf{Definition 2.1.} Let $U$ a force function per unit mass; $(J,E)\in \textbf{R}^{2}$ is an \emph{angular momentum-energy state}
if it exists a motion of the particle with $J$ the angular momentum per unit mass and $E$ the total energy per unit mass. The \emph{Angular
Momentum-Energy Space} $\textbf{S}_{U}$ is the set of the angular momentum-energy states. \vspace{2mm}

\noindent \textbf{Remark 2.2.} $(J,E)\in \textbf{R}^{2}$ is an angular momentum-energy state if and only if it exists a solution $(r,\varphi )$
of (2) and (3). \vspace{2mm}

\noindent \textbf{Remark 2.3.} An angular momentum-energy state $(J,E)$ is corresponding to an uniform rotation if exists an uniform rotation of
the particle with $J$ the angular momentum per unit mass and $E$ the total energy per unit mass. We denote by $\textbf{S}_{U}^{u.r}$ \emph{the
set of the angular momentum-energy states which are corresponding to the uniform rotations}.

\vspace{10mm}

\centerline{\bf 3. The Angular Momentum-Energy Space}

\vspace{5mm}

\textbf{3.1. The set of the angular momentum-energy states which are corresponding to the uniform rotations}. This set is characterized by the
theorem 1, we have:
\begin{equation}\label{*}
\textbf{S}_{U}^{u.r}=\{(J,E)\mid \exists s>0 \,\, J^{2}=s^{3}U'(s)\,\,\hbox{and}\,\, E=U(s)+\frac{sU'(s)}{2}\}
\end{equation}

It is easy to see that we have the following characterizations of the set of angular momentum-energy states which are corresponding to the
uniform rotations:
\begin{equation}\label{*}
\textbf{S}_{U}^{u.r}=\{(J,E)\mid \exists s>0 \,\, (\textbf{V}_{J}^{U})'(s)=0\,\,\hbox{and}\,\, E=\textbf{V}_{J}^{U}(s)\}
\end{equation}
\begin{equation}\label{*}
\textbf{S}_{U}^{u.r}=\{(J,E)\mid \exists s>0 \,\, J^{2}=\textbf{W}_{E}^{U}(s)\,\,\hbox{and}\,\, (\textbf{W}_{E}^{U})'(s)=0 \}
\end{equation}

We present some interesting properties of the set $\textbf{S}_{U}^{u.r}.$
\vspace{2mm}

\noindent \textbf{Proposition 3.1.} i) \emph{If $(J,E)\in \textbf{S}_{U}^{u.r}$, then $(-J,E)\in \textbf{S}_{U}^{u.r}$.}

\noindent ii) \emph{If $r_{0}>0$, then it exists an uniform rotation $r(t)=r_{0}$ if and only if $U^{\prime }(r_{0})\geq 0$.}

\noindent iii) \emph{If ${\min}_{r>0}\textbf{V}_{J}^{U}(r)\in \textbf{R}$ and $E=\min_{r>0}\textbf{V}_{J}^{U}(r)$, then $(J,E)\in
\textbf{S}_{U}^{u.r}$.}

\noindent iv) \emph{If $\max_{r>0}\textbf{W}_{E}^{U}(r)\in \textbf{R}_{+}$ and $J=\sqrt{\max_{r>0}\textbf{W}_{E}^{U}(r)}$ then $(J,E)\in
\textbf{S}_{U}^{u.r}$.} \vspace{2mm}

\noindent \emph{Proof.} The first and second results are consequences of the characterization (7) of the set $\textbf{S}_{U}^{u.r}.$

\noindent iii) In our hypotheses it exists $r_{0}>0$ such that $E=\textbf{V}_{J}^{U}(r_{0}).$ According to Fermat theorem we have
$(\textbf{V}_{J}^{U})^{\prime }(r_{0})=0.$ It is easy to see that the relations (4) are verified and $(J,E)\in \textbf{S}_{U}^{u.r}.$

The proof of iv) is analogue with the demonstration of iii). \vspace{2mm}

\textbf{3.2. The properties of the Angular Momentum-Energy Space}. Firstly we present a characterization of the Angular Momentum-Energy Space
using the properties of the force function per unit mass $U$. \vspace{2mm}

\noindent \textbf{Theorem 3.1.} \emph{The Angular Momentum-Energy Space is characterized by:}
\begin{equation}\label{*}
\textbf{S}_{U}=\{(J,E) \mid E>\inf_{r>0}\textbf{V}_{J}^{U}(r)\,\,\hbox{or}\,\,E=\min_{r>0}\textbf{V}_{J}^{U}(r)\}\
\end{equation} and
\begin{equation}\label{*}
\textbf{S}_{U}=\{(J,E) \mid J^{2}<\sup_{r>0}\textbf{W}_{E}^{U}(r)\,\,\hbox{or}\,\,J^{2}=\max_{r>0}\textbf{W}_{E}^{U}(r)\}\
\end{equation}
\vspace{2mm}

\emph{Proof.} We suppose that $(J,E)\in \textbf{S}_{U}.$ It exists $(r,\varphi )$ a solution of (2) and (3). It is easy to see that $E\geq
\inf_{r>0}\textbf{V}_{J}^{U}(r)$ and $J^{2}\leq\sup_{r>0}\textbf{W}_{E}^{U}(r)$.

If $E= \inf_{r>0}\textbf{V}_{J}^{U}(r)$, then it exists $r_{0}>0$ such that $E=\textbf{V}_{J}^{U}(r_{0})=\min_{r>0}\textbf{V}_{J}^{U}(r).$ We
deduce that:
$$\textbf{S}_{U}\subset\{(J,E) \mid E>\inf_{r>0}\textbf{V}_{J}^{U}(r)\,\,\hbox{or}\,\,E=\min_{r>0}\textbf{V}_{J}^{U}(r)\}$$
If $J^{2}=\sup_{r>0}\textbf{W}_{E}^{U}(r)$, then it exists $r_{0}>0$ such that $J^{2}=\textbf{W}_{E}^{U}(r_0)=\max_{r>0}\textbf{W}_{E}^{U}(r)$.
We deduce that:
$$\textbf{S}_{U}\subset \{(J,E) \mid J^{2}<\sup_{r>0}\textbf{W}_{E}^{U}(r)\,\,\hbox{or}\,\,J^{2}=\max_{r>0}\textbf{W}_{E}^{U}(r)\}$$

Let $(J,E)\in \{(J,E) \mid E>\inf_{r>0}\textbf{V}_{J}^{U}(r)\,\,or\,\,E=\min_{r>0}\textbf{V}_{J}^{U}(r)\}.$ If
$E>\inf_{r>0}\textbf{V}_{J}^{U}(r)$, then it exists $r_{0}>0$ such that $E>\textbf{V}_{J}^{U}(r_{0})$. We consider the Cauchy problem of
differential equations:
$$\dot{\varphi}=\frac{J}{r^{2}},\,\,\,\dot{r}=2\sqrt{E-\textbf{V}_{J}^{U}(r)},\,\,\,\varphi(0)=\frac{J}{r_{0}^{2}},\,\,\,r(0)=r_{0}.$$
According to the Cauchy-Lipschitz Theorem the Cauchy problem has a solution $(r,\varphi )$. This solution is not an uniform rotation and it is a
solution of the system (3). Using the Theorem 2.1 we deduce that $(r,\varphi )$ is a solution of (2) and we conclude that:
$$\textbf{S}_{U}\supset\{(J,E) \mid E>\inf_{r>0}\textbf{V}_{J}^{U}(r)\,\,\hbox{or}\,\,E=\min_{r>0}\textbf{V}_{J}^{U}(r)\}$$
Let $(J,E)\in \{(J,E) \mid J^{2}<\sup_{r>0}\textbf{W}_{E}^{U}(r)\,\,\hbox{or}\,\,J^{2}=\max_{r>0}\textbf{W}_{E}^{U}(r)\}$. If
$J^{2}<\sup_{r>0}\textbf{W}_{E}^{U}(r)$, then it exists $r_{0}>0$ such that $J^{2}<\textbf{W}_{E}^{U}(r_{0})$. We consider the Cauchy problem of
differential equations:
$$\dot{\varphi}=\frac{J}{2u},\,\,\,\dot{u}=\sqrt{\textbf{W}_{E}^{U}(\sqrt{2u})-J^{2}},\,\,\,\varphi(0)=\frac{J}{r_{0}^{2}},\,\,\,u(0)=\frac{r_{0}^{2}}{2}$$
According to the Cauchy-Lipschitz Theorem the Cauchy problem has a solution $(u,\varphi )$. In this situation $(r,\varphi )=(\sqrt{2u},\varphi
)$ is a solution of the system (3). We  obtain:
$$ \textbf{S}_{U}\supset \{(J,E) \mid
J^{2}<\sup_{r>0}\textbf{W}_{E}^{U}(r)\,\,\hbox{or}\,\,J^{2}=\max_{r>0}\textbf{W}_{E}^{U}(r)\}$$

We present some properties of the Angular Momentum-Energy Space.
\vspace{2mm}

\noindent \textbf{Theorem 3.2.} \emph{i) If $(J,E)\in \textbf{S}_{U}$, then $(-J,E)\in \textbf{S}_{U}.$}

\noindent \emph{ii) If $k\in \textbf{R}$, then $\textbf{S}_{U+k}=\textbf{S}_{U}+(0,k).$}

\noindent \emph{iii) Let $U_{1},U_{2}$ two force functions, if $U_{1}\leq U_{2}$, then $\textbf{S}_{U_{2}}\subset \textbf{S}_{U_{1}}.$}
\vspace{2mm}

\noindent \emph{Proof.} i) We observe that $\textbf{V}_{J}^{U}=\textbf{V}_{-J}^{U}$ and one obtains the affirmation.

\noindent ii) The result is an immediate consequence of the relation $\textbf{V}_{J}^{U+k}=\textbf{V}_{J}^{U}+k.$

\noindent iii) We have $\textbf{V}_{J}^{U_{1}}(r)\leq \textbf{V}_{J}^{U_{2}}(r)$ $\forall r>0$ which implies easily our proposition.
\vspace{2mm}

\noindent \textbf{Remark 3.1.} $U_{1}\leq U_{2} \Leftrightarrow \forall r>0$ we have $U_{1}(r)\leq U_{2}(r)$.

\vspace{2mm}

Finally we study the conditions of the force function per unit mass $U$ such that the Angular Momentum-Energy Space is the entire
$\textbf{R}^{2}.$ \vspace{2mm}

\noindent \textbf{Theorem 3.3.} \emph{The next affirmations are equivalents:}

\noindent i) $\textbf{S}_{U}=\textbf{R}^{2}.$

\noindent ii) $\liminf_{r\rightarrow 0}r^{2}U(r)=-\infty $ or $\liminf_{r\rightarrow \infty }U(r)=-\infty .$ \vspace{2mm}

\noindent \emph{Proof.} Firstly we suppose that $\textbf{S}_{U}=\textbf{R}^{2}$, $\liminf_{r\rightarrow 0}r^{2}U(r)>-\infty $ and
$\liminf_{r\rightarrow \infty }U(r)>-\infty $.

It exists $0<r^{\ast }<r^{\ast \ast }$ and $k^{\ast },k^{\ast \ast }\in \textbf{R}_{+}^{\ast }$ such that, if $r\in (0,r^{\ast })$, then
$U(r)>-\frac{k^{\ast }}{r^{2}}$ and if $r>r^{\ast \ast }$, then $U(r)>-k^{\ast \ast }.$ $U$ is a continuous function and $[r^{\ast },r^{\ast
\ast }]$ is a compact interval, there exists $k^{\ast \ast \ast }>0$ such that $U(r)>k^{\ast \ast \ast }$ for all $r\in \lbrack r^{\ast
},r^{\ast \ast }].$ We introduce $\widetilde{k}=\max \{k^{\ast },k^{\ast \ast \ast }r^{\ast 2},k^{\ast \ast }r^{\ast \ast 2}\}>0$ and we have
$U(r)\geq -\frac{\widetilde{k}}{r^{2}}$ for all $r>0.$ Using the Theorem 3.2. one obtains that $\textbf{S}_{U}\subset
\textbf{S}_{-\frac{\widetilde{k}}{r^{2}}}.$ We known that $\textbf{S}_{-\frac{\widetilde{k}}{r^{2}}}\neq \textbf{R}^{2}$ (see the \S 4.3.) and
we deduce that $\textbf{S}_{U}\neq \textbf{R}^{2},$ but this result is a contradiction.

If the affirmation ii) is true, then $\inf_{r>0}\textbf{V}_{J}^{U}(r)=-\infty $ for all $J\in \textbf{R}.$ According to Theorem 3.1. we obtain
that the affirmation i) is true.

\newpage

\centerline{\bf 4. Particular cases of Angular Momentum-Energy Space}

\vspace{5mm}

\noindent \textbf{4.1. An Isolated Particle; $U=0$}.
\begin{equation}\label{*}
\textbf{S}_{0}=\{(J,E)\in \textbf{R}^{2}\,/\,E>0\}\cup \{(0,0)\},\,\,\,\,\textbf{S}_{0}^{u.r}=\{(0,0)\}
\end{equation}

In this case an uniform rotation is an equilibrium point. The angular momentum-energy state $(0,0)$ is corresponding to all uniform rotations
(equilibrium points).
\vspace{2mm}

\noindent \textbf{Remark 4.1.} Let $k\in R.$ Using the theorem 3 we obtain:
\begin{equation}\label{*}
\textbf{S}_{k}=\{(J,E)\,/\,E>k\}\cup \{(0,0)\}
\end{equation}
\vspace{2mm}

\noindent \textbf{4.2. Particle in a Gravitational Force Field, $U=-\frac{k}{r}$}. We suppose that the gravitational force is an attraction
force ($k>0).$ In this case we have:
\begin{equation}\label{*}
\textbf{S}_{-\frac{k}{r}}=\{(J,E)\,/\,EJ^{2}\geq
-\frac{k^{2}}{2}\},\,\,\,\,\textbf{S}_{-\frac{k}{r}}^{u.r}=\{(J,E)\,/\,EJ^{2}=-\frac{k^{2}}{2}\}
\end{equation}
\vspace{2mm}

\noindent \textbf{4.3. $U=-\frac{k}{r^{2}}$ with $k>0$}. We have:
\begin{equation}\label{*}
\textbf{S}_{-\frac{k}{r^{2}}}=S_1\bigcup S_2\bigcup S_3,\,\,\,\,\textbf{S}_{-\frac{k}{r^{2}}}^{u.r}=\{(-\sqrt{2k},0),(\sqrt{2k},0)\}
\end{equation}
where:
$$\left\{%
\begin{array}{ll}
    S_1=\{(J,E)\,/\,(J^{2}>2k\,\,\,\hbox{and}\,\,\, E>0)\} \\
    S_2=\{(J,E)\,/\,J^{2}=2k\,\,\,\hbox{and}\,\,\, E\geq 0\} \\
    S_3=\{(J,E)\,/\,J^{2}<2k\,\,\,\hbox{and}\,\,\, E\in \textbf{R}\} \\
\end{array}%
\right.$$
 \vspace{2mm}

\noindent \textbf{4.4. Particle in a Hooke Force Field, $U=\frac{k}{2}r^{2}$ with $k>0$}. We have:
\begin{equation}\label{*}
\textbf{S}_{\frac{k}{2}r^{2}}=\{(J,E)\,/\,E\geq
\sqrt{k}|J|\}-\{(0,0)\},\,\,\,\,\textbf{S}_{\frac{k}{2}r^{2}}^{u.r}=\{(J,E)\,/\,E=\sqrt{k}|J|\}-\{(0,0)\}
\end{equation}
\vspace{2mm}

\noindent \textbf{Remark 4.2.} For us $U$ is not defined for $r=0; $ this is the reason for which $(0,0)$ is not a angular momentum-energy
state. \vspace{2mm}

\noindent \textbf{4.5. Particle in a Repulsive Elastic force Field, $U=-\frac{k}{2}r^{2}$ with $k>0$}. In this case:
\begin{equation}\label{*}
\textbf{S}_{-\frac{k}{2}r^{2}}=\textbf{R}^{2},\,\,\,\,\textbf{S}_{-\frac{k}{2}r^{2}}^{u.r}=\emptyset
\end{equation}
\vspace{2mm}

\noindent \textbf{4.6. $U=-\frac{k}{r}-\frac{q}{r^{2}}$ with $k>0$ and $q>0$}.
\begin{equation}\label{*}
\textbf{S}_{-\frac{k}{r}-\frac{q}{r^{2}}}=\{(J,E)\,/\,J^{2}\leq 2q\,\,or\,\,(J^{2}>2q\,\,\hbox{and}\,\, E(J^{2}-2q)\geq -\frac{k^{2}}{2})\}
\end{equation}
\begin{equation}\label{*}
\textbf{S}_{-\frac{k}{r}-\frac{q}{r^{2}}}^{u.r}=\{(J,E)\,/ E<0\,\,\hbox{and}\,\, E(J^{2}-2q)=-\frac{k^{2}}{2}\}
\end{equation}
\vspace{2mm}

\noindent \textbf{4.7. $U=-\frac{k}{r^{2n}}$ with $k>0$ and $n>0$}. If $n>1$, then we have:
\begin{equation}\label{*}
\textbf{S}_{-\frac{k}{r^{2n}}}=\textbf{R}^{2},\,\,\,\,\textbf{S}_{-\frac{k}{r^{2n}}}^{u.r}=\{J,E)\,/\,EJ^{\frac{2n}{1-n}}=(n-1)(2n)^{\frac{n}{1-n}}k^{\frac{1}{1-n}}\}
\end{equation}
The case $n=1$ is studied in the \S 4.3.

If $n\in (0,1)$ then:
\begin{equation}\label{*}
\textbf{S}_{-\frac{k}{r^{2n}}}=\{J,E)\,/\,EJ^{\frac{2n}{1-n}}\geq -(1-n)(2n)^{\frac{n}{1-n}}k^{\frac{1}{1-n}}\}
\end{equation}
\begin{equation}\label{*}
\textbf{S}_{-\frac{k}{r^{2n}}}^{u.r}=\{J,E)\,/\,EJ^{\frac{2n}{1-n}}=-(1-n)(2n)^{\frac{n}{1-n}}k^{\frac{1}{1-n}}\}
\end{equation}
\vspace{2mm}

\noindent \textbf{4.8. $U=q\sin \frac{1}{r}$ with $q>0$}. This case is interesting for theoretical reasons. Using the theorem of
characterization of the angular momentum-energy Space we obtain:
\begin{equation}\label{*}
\textbf{S}_{q\sin \frac{1}{r}}=\{(J,E)\,/\,E>-q\}\cup \{(0,-q)\}
\end{equation}

The set $D_{q\sin \frac{1}{r}}$ of the distances $r_{0}$ for which exists an uniform rotations with $r(t)=r_{0}$ is described by the formula:
\begin{equation}\label{*}
D_{q\sin \frac{1}{r}}=\cup_{k\in \textbf{N}}[\frac{2}{(4k+3)\pi },\frac{2}{(4k+1)\pi }]
\end{equation}

If $r_{0}\in \{\frac{2}{(2k+1)\pi }\,/\,k\in \textbf{N}\}$, then at the distance $r_{0}$ the particle can have an equilibrium state. All
equilibrium states have an angular momentum-energy state in the set $\{(0,q),(0,-q)\}.$

The set of the angular momentum-energy states which are corresponding to the uniform rotations is:
\begin{equation}\label{*}
\textbf{S}_{q\sin \frac{1}{r}}^{u.r}=\{(J,E)\,/\,\exists s>0\,\,\,J^{2}=-qs\cos \frac{1}{s}\,\,\,\hbox{and}\,\,\,E=q\sin
\frac{1}{s}-\frac{q}{2s}\cos \frac{1}{s}\}
\end{equation}

\vspace{10mm}

\centerline{\bf References}

\vspace{5mm}

{\footnotesize

\begin{list}{{[\rm\arabic{obs}]}}
{
\usecounter{obs}
\setlength{\rightmargin}{0mm} }

\item Arnold V.I., Mathematical Methods of Classical Mechanics, Springer Verlag, 1989.

\item Chon H., Kulsrud R.M., \emph{The stellar distribution around a black hole: numerical integration of the Fokker-Plank equation}, The
Astrophysical Journal, 226: 1087-1108, 1978.

\item Yi Y.G., \emph{General-relativistic equations of motion in terms of energy and angular momentum}, European Journal of Physics, 2003, vol.
24, no.4, pp. 413-417.

\item Com\u anescu D., Modele \c{s}i metode \^{i}n mecanica punctului material, Mirton Publishing House, Timi\c{s}oara, 2004.

\item Landau L.D., Lifsit E.M., Mechanics, Ed. Tehnica, Bucure\c{s}ti, 1966.

\end{list}

\vspace{10mm}

\noindent Authors' affiliation: Department of Mathematics, Faculty
of Mathematics and Computer Science, West University of Timi\c
soara, Romania

\vspace{2mm}

\noindent E-mail: comanescu@math.uvt.ro

\end{document}